\documentclass{JHEP3}

\usepackage{epsfig}

\newcommand\fverb{\setbox\pippobox=\hbox\bgroup\verb}
\newcommand\fverbdo{\egroup\medskip\noindent%
                        \Fbox{\unhbox\pippobox}\ }
\newcommand\fverbit{\egroup\item[\fbox{\unhbox\pippobox}]}
\newbox\pippobox
\def\JHEP{\it JHEP}

\newcommand{\newc}{\newcommand}
\newc\eg{{\it {e.g.}}} \newc\etal{{\it {et al.}}} \newc\ie{{\it i.e.}}
\newc\etc{{\it {etc}}} 
\newcommand\fa{f_{a}}
\newcommand\mchi{m_{\chi}}

\newcommand\axino{\widetilde{a}}        
\newcommand\gluino{\widetilde{g}}
\newcommand\squark{\widetilde{q}}

\newcommand\neut{\chi}

\newcommand\maxino{m_{\tilde a}}
\newcommand\treh{T_R}
\newcommand\abunda{\Omega_{\tilde a}h^2}
\newc{\cachigamma}{C_{a\neut\gamma}}
\newc{\caww}{C_{aWW}}                   
\newc{\cayy}{C_{aYY}}
\newc{\sthw}{\sin\theta_W}              
\newc{\cthw}{\cos\theta_W}
\newc{\bino}{\widetilde B}              
\newc{\wino}{\widetilde W_3}
\newc{\higgsinob}{{\widetilde H}^0_b}   
\newc{\higgsinot}{{\widetilde H}^0_t}
\newc{\abund}{\Omega h^2}
\newc{\abundchi}{\Omega_\neut h^2}
\newc{\rhocrit}{\rho_{crit}}
\newc{\rhochi}{\rho_{\neut}}
\newc{\mplanck}{M_{\rm P}}
\newc{\xf}{x_f}
\newc{\jxf}{J({\xf})}
\newc{\VEV}[1]{\langle #1 \rangle}
\newcommand\tev{\,\mbox{TeV}}
\newcommand\gev{\,\mbox{GeV}}

\newcommand\ev{\,\mbox{eV}}

\newc\gbar{{\overline{g}}}

\newc{\ra}{\rightarrow}

\newc{\beq}{\begin{equation}}
\newc{\eeq}{\end{equation}}
\newc{\bea}{\begin{eqnarray}}
\newc{\eea}{\end{eqnarray}}
\renewcommand\({\left(}
\renewcommand\){\right)}
\renewcommand\[{\left[}
\renewcommand\]{\right]}

\newc{\nspin}{n_{\rm spin}}
\newc{\nflavor}{n_{\rm F}}
\newc\vrel{v_{\rm rel}}

                \newcommand\omegaatp{\Omega^{\rm TP}_{\tilde a}}
\newc{\naxino}{n_{\tilde a}}
\newc{\ngamma}{n_\gamma}
\newc{\ychi}{Y_{\chi}}                  \newc{\yeqchi}{Y^{\rm EQ}_{\chi}}
\newc{\yaxino}{Y_{\tilde a}}
\newc{\yeqaxino}{Y^{\rm EQ}_{\tilde a}}
\newc{\ythaxino}{Y^{\rm TP}_{\tilde a}}
\newc{\ynthaxino}{Y^{\rm NTP}_{\tilde a}}
\newc{\yascat}{Y^{\rm scat}_{i,j}}      \newc{\yadec}{Y^{\rm dec}_{i}}
\newc{\gstar}{g_\ast}           \newc{\gsstar}{g_{s\ast}}

\def\lag             {{\cal L}}

\newcommand\lsim{\mathrel{\rlap{\lower4pt\hbox{\hskip1pt$\sim$}}
    \raise1pt\hbox{$<$}}}
\newcommand\gsim{\mathrel{\rlap{\lower4pt\hbox{\hskip1pt$\sim$}}
    \raise1pt\hbox{$>$}}}

       \def\pslash{\not{\hbox{\kern-2.3pt $p$}}}
       \def\kslash{\not{\hbox{\kern-2.3pt $k$}}}
       \def\qslash{\not{\hbox{\kern-2.3pt $q$}}}
       \def\ddslash{\not{\hbox{\kern-2.3pt $d$}}}

\title{Effects of Squark Processes on the \\ Axino CDM Abundance}

\author{Laura Covi \\
    DESY Theory Group,
        Notkestrasse 85, 22603 Hamburg, Germany \\
        E-mail: \email{Laura.Covi@desy.de}}

\author{Leszek Roszkowski \\
        Department of Physics, 
        Lancaster University, Lancaster LA1 4YB, England\\
        E-mail: \email{L.Roszkowski@lancaster.ac.uk}}

\author{Michael Small \\
        Department of Physics, 
        Lancaster University, Lancaster LA1 4YB, England\\
        E-mail: \email{m.small@lancaster.ac.uk}}

\abstract{
We investigate the role of an effective dimension-$4$
axino-quark-squark coupling in the thermal processes producing stable
cold axino relics in the early Universe.  We find that, while the
induced squark and quark scattering processes are always negligible,
squark decays become important in the case of low reheat temperature
and large gluino mass. The effect can tighten the bounds on the
scenario from the requirement that cold dark matter axinos do not
overclose the Universe.
}

\keywords{Supersymmetric Effective Theories, Cosmology of Theories beyond 
            the SM, Dark Matter}

\preprint{DESY 02-080}    % OR: \preprint{Aaaa/Mm/Yy\\Aaa-aa/Nnnnnn}
\begin{document}

% \maketitle %%%%%%%%%% THIS IS IGNORED %%%%%%%%%%%

%%%%%%%%%%%%%%%%%%%%%%%%%%%%%%%%%%%%%%%%%%%%%%%%%%%%%%%%%%%%%%%%%%%%%%%%%%%%
\section{Introduction}\label{sec:intro}

The nature of the dark matter (DM) in the Universe remains one of the
most challenging problems in cosmology. Numerous candidates for DM
have been proposed in the literature. One of the most popular
candidates in the context of supersymmetric theories with $R$-parity
conservation is the lightest supersymmetric particle (LSP). Such a
candidate has usually been studied in the framework of supersymmetric
extensions of the Standard Model, such as the Minimal Supersymmetric
Standard Model (MSSM)~\cite{MSSM-dm} or its constrained version
(CMSSM)~\cite{CMSSM-dm}.  In these models, the LSP is often the
lightest neutralino, \ie~a linear combination of the fermionic
superpartners of neutral gauge and Higgs bosons.  The interactions of
the neutralino are weak, and its number density at decoupling is
therefore often of the required order of magnitude, which makes
it an excellent candidate for the WIMP (Weakly Interacting
Massive Particle)~\cite{wimp}.

The case of the neutralino WIMP is very appealing and puts strong
bounds on SUSY masses and 
parameters~\cite{MSSM-dm,CMSSM-dm}, but it is certainly not a unique one.
One should not forget that very 
different possibilities can arise in other well-motivated
extensions of the Standard Model.
For example, in the framework of supergravity there arises the
gravitino, the SUSY partner of the graviton. The gravitino has long
been known to be a potential candidate for
DM~\cite{gravitinoproduction}.  A number of more speculative
possibilities, stemming from, for example, string theories, have also
been proposed.

The aim of this paper is to investigate another well-motivated
scenario, in which the LSP is the axino. Such a particle is present in
SUSY models that incorporate the Peccei-Quinn (PQ) solution to the
strong CP problem. The PQ global $U(1)$ symmetry~\cite{pq} was
introduced more than 20 years ago to solve the strong CP problem, and
still remains the most elegant solution, especially since lattice
results seem to disfavor the possibility of a massless
up-quark~\cite{lattice}.  The axino does not belong to the usual MSSM
spectrum, but, as a fermionic superpartner of the
axion~\cite{axion,axionreviews:cite}, is much more weakly
interacting. Similarly to the case of the axion, the couplings of the
axino to ``ordinary'' matter (below simply referred to as ``matter'')
are suppressed by the Peccei-Quinn scale $\fa\sim10^{11}\gev$, thus
the axino remains invisible to present experimental searches.

A stable relic axino has been shown to be an attractive candidate for
either warm~\cite{kmn,rtw,bgm,ckkr} or, preferably, cold dark matter
(CDM)~\cite{ckkr,ckr}. An underlying assumption is that a primordial
axino population was washed away by an earlier inflationary
period. Axinos can then again be produced in large numbers through
several possible mechanisms. One efficient way is through {\em thermal
production} (TP); this proceeds via thermal scatterings involving, \eg,
gluinos (and their decays) in the primordial plasma~\cite{rtw,ckkr},
in close analogy with gravitino
production~\cite{gravitinoproduction}. Another one is {\em non-thermal
production} (NTP): an example of this being an out-of-equilibrium
decay of the next-to-lightest particle (NLSP)~\cite{kmn,ckr}.  A
number of papers have investigated cosmological properties of axinos
as warm~\cite{kmn, bgm, ay00} or cold relics~\cite{ckl00,kk02}.

In a recent extensive analysis~\cite{ckkr}, the role of
the effective dimension-5 gluino-gluon-axino interaction was examined
in the production of stable axinos as relics in the Universe. At
high reheat temperatures $T_R$ (much larger than squark/gluino masses
but still $T_R\ll\fa$ in order not to restore the PQ symmetry), the 
thermal production of axinos via scattering processes was shown to be
dominant. This allows one to place strong bounds on the reheat
temperature as a function of the axino mass by imposing the constraint
$\Omega_{\tilde a} h^2 \lsim 1$. At lower $T_R$, gluino decays in the
thermal plasma and, at $T_R\ll m_{\tilde g}$, non-thermal production of
axinos were shown to become more important.

The scope of this paper is to supplement the analysis
in~\cite{ckr,ckkr} by investigating the thermal production of
axinos involving the dimension-4 squark-quark-axino interaction. The effective
vertex for this interaction arises at a one loop order in an effective
theory but we find it to be enhanced by the gluino mass.  For this
reason, we argue that squark decays cannot be neglected in a
computation of the axino abundance, especially in the case of a heavy
gluino.

The paper is organized as follows. In Sec.~\ref{sec:axionmodels} we
briefly review axion models. In Sec.~\ref{sec:ksvzaxino} we
introduce the effective squark-quark-axino vertex which is then
computed in Sec.~\ref{sec:sqa}. The formalism for the thermal
production of axinos is reviewed in Sec.~\ref{sec:thermalaxino} and
applied to two- and three-body squark decays in
Sec.~\ref{sec:sqdecay} and to scattering of quarks and squarks in
Sec.~\ref{sec:dim4}. We present our numerical analysis in
Sec.~\ref{sec:results} and conclude in Sec.~\ref{sec:concl}.

%%%%%%%%%%%%%%%%%%%%%%%%%%%%%%%%%%%%%%%%%%%%%%%%%%%%%%%%%%%%
\section{Axion models}\label{sec:axionmodels}

In the PQ scenario, a complex scalar field is introduced, which breaks
the global $U(1)_{PQ}$ at a high scale and whose Goldstone boson
component, the axion, plays the role of a dynamical $\theta_{QCD}$
relaxing at the origin after the QCD phase transition. In fact, after
the chiral symmetry breaking, the axion acquires a tiny mass from
instanton effects~\cite{qcdanomaly:cite}.

Two classes of models implementing the PQ mechanism are usually studied
in the literature: the KSVZ-type (hadronic) axion models
(KSVZ)~\cite{ksvz} and the DFSZ-type models~\cite{dfsz}. In the first
case, the SM particles are not charged under $U(1)_{PQ}$, while in the
second they are. For this reason, different axion interactions are
present in the two approaches.

In supersymmetric extensions of axion models the axion field is
promoted to a chiral multiplet~\cite{susyaxion} which contains not
only the pseudoscalar axion, but also a scalar, the saxion, and their
fermionic superpartner, the axino. An interesting feature of such
supersymmetric models is that the PQ global symmetry is enlarged to a
complex $U(1)$, which ensures that the whole axion multiplet is
massless and degenerate~\cite{ckn}, as long as supersymmetry and the
chiral symmetry are unbroken.  If supersymmetry is broken at low
energies, as is required to explain the large hierarchy between the
electroweak and the Planck scale, we can assume that this degeneracy
is not strongly lifted, and consider a model in which the axion
multiplet is the only additional low-energy degree of freedom compared
to those already contained in the MSSM.

The breaking of supersymmetry reduces the complex $U(1)$ to the usual
$U(1)_{PQ}$ and generates a soft mass for the saxion~\cite{saxion},
while the axion remains protected by the surviving $U(1)_{PQ}$. For
the axino different scenarios can arise, strongly dependent on the
model and the particle content \cite{ckn,cl}. In general, the axino, being a 
component of a
chiral multiplet, does not acquire a tree-level Majorana soft mass
term. However, a Dirac mass term may arise which will lead to mixing
with the neutral heavy states in the PQ sector and the MSSM neutral
higgsinos, giving rise to an enlarged neutralino mass matrix at low
energy.

On the other hand, if the mixing of the neutralino-axino is very tiny,
or absent, an axino nearly massless mass eigenstate survives at the
tree level (see~\cite{ckn,cl,ckkr}).  Particularly interesting is the
case in which a Majorana mass for the axino is generated at one
loop~\cite{Moxhay-Yamamoto,Goto-Yamaguchi}, falling naturally in the
range of a few tens of GeV, often below the present bounds on the
neutralino. This bound obviously does not apply to an axino state due
to its suppressed interactions with SM particles.

Phenomenologically, an appealing scenario arises: the LSP axino
remains unobserved due to the very tiny couplings of the axion
multiplet with matter, and does not spoil the precision
measurements of the SM, but still plays a key role in the cosmological
evolution of the Universe.

%%%%%%%%%%%%%%%%%%%%%%%%%%%%%%%%%%%%%%%%%%%%%%%%%%%%%%%%%%%%%%%%%%%%%%
\section{The interactions of the KSVZ axino}\label{sec:ksvzaxino}

In this paper we consider mainly an axino of the KSVZ type; 
our final results are expected to be similar to the case
of the  DFSZ axino, and we will comment on the differences below.

In the KSVZ, or hadronic, axion models, the axion multiplet
(and therefore also the axino) does not couple directly with ordinary
matter, instead its interactions with matter are  suppressed not only by
the PQ scale, but also by loop factors. 

A simple example for the superpotential for a KSVZ axion multiplet 
is~\cite{jekim83}
\beq
W= \lambda \Phi Q\bar Q + S (\fa^2 - \Phi \bar \Phi )
\label{W_KSVZ}
\eeq
where $Q, \bar Q$ are PQ-charged, heavy colored triplet states, 
$\Phi$ and $\bar \Phi$ are PQ-charged SM singlets, while $S$ is a 
singlet with respect to both the SM and PQ symmetries. 

The axion multiplet is in this simple case contained in $\Phi$ and 
$ \bar \Phi$: writing their scalar and fermionic part in terms
of the axion multiplet ${\cal A} = (A, \tilde a)$ and another 
complex scalar and fermion pair $(B, \psi)$, we obtain respectively
\bea
\Phi \rightarrow \left( \fa \exp\left[(B+A)/(\sqrt{2}\fa)\right], 
(\psi + \tilde a)/\sqrt{2} \right) \nonumber\\
\bar \Phi \rightarrow \left(
\fa \exp\left[(B-A)/(\sqrt{2}\fa)\right], (\psi- \tilde a)/\sqrt{2}
\right).
\eea

From the superpotential (\ref{W_KSVZ}), we see that in the supersymmetric
vacuum with $\langle \Phi \bar\Phi \rangle = \fa^2 $, 
only the axion multiplet remains massless and all the 
other states acquire a mass of the order of $\fa$.
Note also that no direct coupling between the axion
and the MSSM fields is present in~(\ref{W_KSVZ}). However,  
axino couplings to MSSM fields do arise in the low energy effective
action and will be derived below.

%%%%%%%%%%%%%%%%%%%%%%%%%%%%%%%%%
\EPSFIGURE{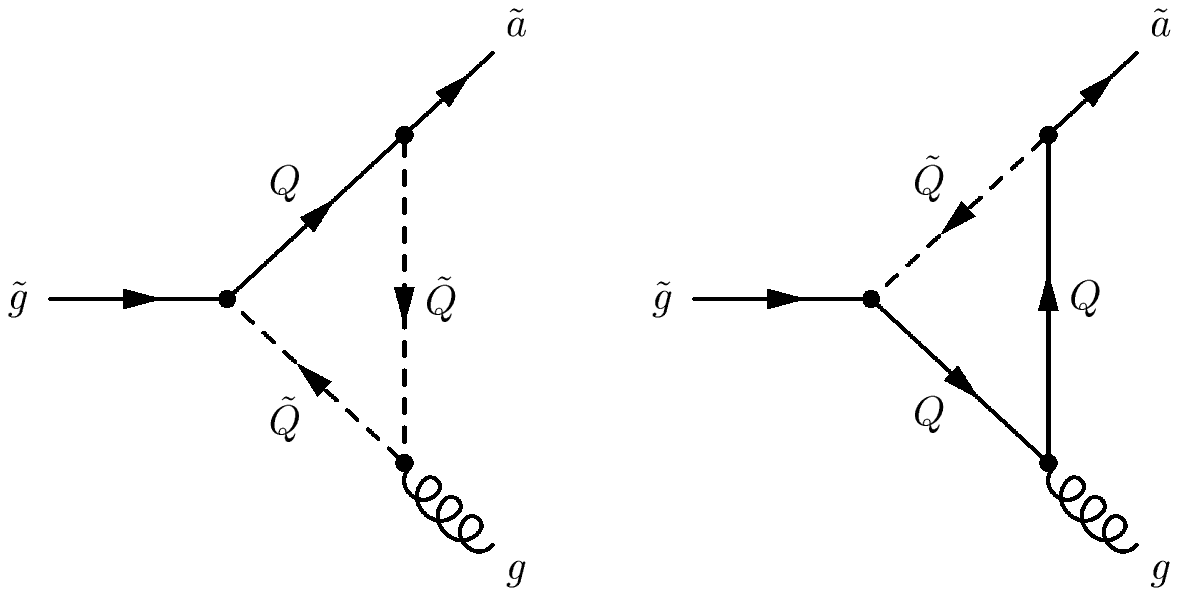, height=2.3in}      
{The two Feynman diagrams contributing to the gluino-gluon-axino 
    interaction. The extra quark $Q$ and squark
    $\widetilde Q$ states are all very heavy, $m_{Q,\widetilde
    Q}\sim\fa$, which allows one to contract the loops to an effective
    ``tree-level'' gluino-gluon-axino vertex. 
\label{fig:loop0}
}
%%%%%%%%%%%%%%%%%%%%%%%%%%%%%%%%

The interaction of the axino with gluons and gluinos proceeds via
the anomaly one-loop diagrams shown in Fig.~\ref{fig:loop0}. 
After integrating out the heavy KSVZ quarks $Q$ and squarks
$\tilde{Q}$ ($m_{Q,\widetilde Q}\sim\fa$), 
we are left with a single effective gluino-gluon-axino dimension-5 
interaction term in the Lagrangian
\beq
\label{eq:agg}
\lag_{\tilde a g \tilde g}= \frac{\alpha_s}{8 \pi (\fa/N)}
\bar{\axino} \gamma_5 \sigma^{\mu\nu} \gluino^b G^b_{\mu\nu}
\eeq
where $b=1,\ldots,8$, $\gluino$ is the gluino and $G$ is the strength of
the gluon field, while 
$N$ is the number of flavors of quarks with the Peccei-Quinn
charge, and is equal to $1$ ($6$) in KSVZ (DFSZ) models. For the remainder of
this paper, we understand $\fa$ to mean $\fa/N$.

The coupling~(\ref{eq:agg}) can also be obtained 
by supersymmetrizing the axion-gluon-gluon interaction, 
which corresponds to adding the following Wess-Zumino term to the MSSM 
superpotential:
\beq
W_{WZ} = {\alpha_s \over 2 \sqrt{2} \pi \fa} {\cal A} \;
Tr \left[ W_\alpha W^\alpha \right],
\eeq
where ${\cal A}$ is the axion multiplet, $W_\alpha $ the vector
multiplet containing the gluon, and the trace sums over color indices.
An interesting characteristic is that the coupling above
is only determined by the QCD anomaly of the heavy states
and is not subject to renormalization~\cite{anom-ren} 
or model dependence.

In an analogous way, Wess-Zumino terms can arise also for the 
other gauge interactions,
depending on the charge of the heavy quark multiplet; at high energy,
when all leptons can be considered massless, such interactions can be 
rotated into the $U(1)_Y$ direction~\cite{ckkr} and we are left to
consider only the term
\beq
W_{WZ} = {\alpha_Y \cayy \over 4 \sqrt{2} \pi \fa} {\cal A} \;
 B_\alpha B^\alpha \, ,
\eeq
where  $B_\alpha $ is the hypercharge vector multiplet and
$ \cayy $ is a model-dependent factor~\cite{coupling}, which vanishes 
if the heavy KSVZ quarks are electrically neutral. If 
they have electric charge $e_Q=-1/3,+2/3$, then $C_{aYY}=2/3,8/3$.
This superpotential term leads to the addition of the following 
effective dimension-5 interaction term to the low-energy Lagrangian
\beq
\label{eq:abb}
\lag_{\tilde a B \tilde B}= \frac{\alpha_Y C_{aYY}}{8 \pi \fa}
\bar{\axino} \gamma_5 \sigma^{\mu\nu} \tilde B B_{\mu\nu}
\eeq
Cosmological implications of the effective axino-gauge boson-gaugino 
operators~(\ref{eq:agg}) and ~(\ref{eq:abb})
have been extensively studied in~\cite{ckkr}. 

However, in addition there arises the effective dimension-4
coupling of the axino to quarks and squarks when we integrate out the
heavy states:
\beq
\lag_{\tilde a q \tilde q}= 
g_{eff}^{L/R}\; \squark^{L/R}_j\, \bar q_j P_{R/L} \gamma^5 \axino.
\label{eq:sqqa}
\eeq
We find that this effective coupling, which arises at two-loop level
in KSVZ models (and therefore at a one-loop level in the effective
theory valid much below $\fa$), can play an important role in the
thermal production of axinos.

Instead of undertaking the complicated task of a full computation of $
g_{eff}^{L/R} $ in the renormalizable theory, which would give a
result strongly dependent on the unknown parameters of the axion
superpotential anyway, our strategy is to resort to the expedient of
estimating the coupling via an effective theory, given by the MSSM
plus the interaction in Eq.~(\ref{eq:agg}). That is to say, we compute
the two one-loop diagrams in Fig.~\ref{fig:sqqa}, and use our result
to define an effective squark-quark-axino vertex in the low energy
Lagrangian. We are then able to study the effects of the coupling on the 
axino abundance.

%%%%%%%%%%%%%%%%%%%%%%%%%%%%%%%%%
\EPSFIGURE{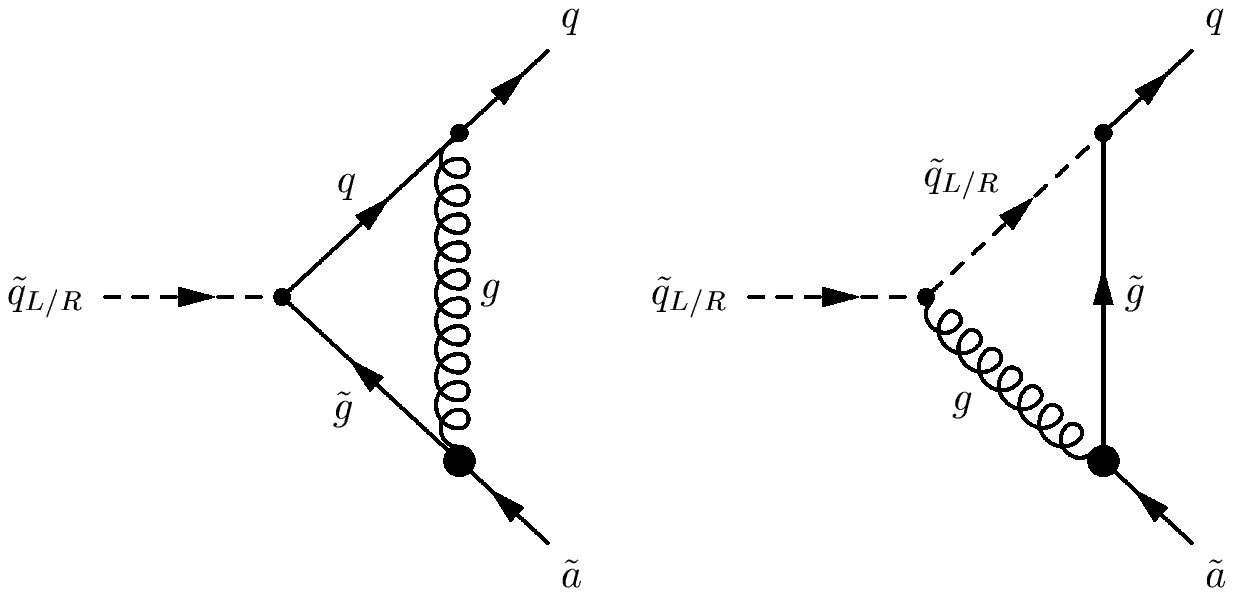, height=2.3in}      
{The Feynman diagrams contributing to the squark-quark-axino 
 interaction. The thick dot denotes the effective 
 gluino-gluon-axino vertex of Fig.~\protect{\ref{fig:loop0}}.
\label{fig:sqqa}
}
%%%%%%%%%%%%%%%%%%%%%%%%%%%%%%%%

Let us now mention the situation for the DFSZ models.
In that case, the axino can mix with all the other neutralinos
through the coupling which generates the Higgs $\mu $ term,
and due to this mixing the coupling in Eq.~(\ref{eq:sqqa}) 
arises at tree level. 
Since the mixing is strongly suppressed by at least the ratio 
of the electroweak to the PQ scale, $ v_{EW}/\fa$, and since we consider the
case when an axino-like light state remains in the spectrum,
we do not expect the full DFSZ result for $g_{eff}^{L/R} $
to be much larger than our result for the KSVZ case.

In general, the effective coupling to a quark and a squark could be enhanced 
for the DFSZ axino, but is not expected to be suppressed, unless there 
are very unlikely cancellations between diagrams. Therefore, the upper bounds 
on the reheat temperature
we obtain for the KSVZ case are also conservative bounds 
for the DFSZ model.

%%%%%%%%%%%%%%%%%%%%%%%%%%%%%%%%%%%%%%%%%%%%%%%%%%%%%%%%%%%%%%%
\section{The effective squark-quark-axino vertex}\label{sec:sqa}

We now proceed to compute the effective dimension-4 squark-quark-axino
interaction defined by the two diagrams in Fig.~\ref{fig:sqqa}. 
The corresponding amplitudes for an axino
$\axino$ with momentum $k_1$, a quark $q$ with momentum $k_2$ and a
squark $\squark^{L/R}$ with momentum $k_1+k_2$, are given by: 
\bea
{\cal A}_1^{L/R} &=& \pm {4 \sqrt{2} \over 3} {\alpha^2_s \over \fa}
\,\, \bar u_{q} (k_2) I_1^{L/R} \gamma^5 v_{\tilde a}(k_1)
\\
{\cal A}_2^{L/R} &=& \mp {8 \sqrt{2} \over 3} {\alpha^2_s \over \fa}
\,\, \bar u_{q} (k_2) I_2^{L/R} \gamma^5 v_{\tilde a}(k_1)
\eea
where summation over color has been performed, 
the upper/lower signs refer to left- and right-handed (s)quarks 
respectively, and $I^{L/R}$'s come from loop integrations. 
We have suppressed the flavor index since the amplitudes are diagonal
in flavor space.

The integral of $I^{L/R}_1$ is logarithmically divergent;
in the full renormalizable KSVZ model (with broken supersymmetry), 
such a divergence would be taken care of
by a counterterm and reabsorbed in the parameters of the 
Lagrangian.
Rather than perform the full renormalization, we instead keep the 
dependence on the UV cut-off $\fa $. 
This procedure is justified by the fact 
that we are working in the context of a non-renormalizable effective theory, 
valid only below $\fa$. At higher scales, all the heavy 
degrees of freedom become dynamical.
With this method, we are absorbing all our ignorance about the high energy 
theory into one single parameter, the ``effective'' PQ scale $\fa$. This 
allows us to perform a model-independent analysis of the implications of the
effective coupling.

We have performed the loop integration using the usual methods,
based on the Feynman parameterization. To lowest order in $1/\fa $,
for $ m_{q_j} \rightarrow 0$ and assuming that the axino mass
is much smaller than the gluino mass $m_{\tilde g}$, we obtain
\bea
I_1^{L/R}  
&=& 
{-3 i m_{\tilde g} \over 8 \pi^2} 
P_{R/L} \[\log\({\fa\over m_{\tilde g}}\) + {1\over 3}
- {1\over 3} \(1-{m^2_{\tilde g}\over 2 k_1 \cdot k_2}\)
\log \(1- {2 k_1\cdot k_2 \over m^2_{\tilde g}}\)\] \\
I_2^{L/R} 
&=&
{- 3 i m_{\tilde g} \over 8 \pi^2} 
P_{R/L} \[-{1\over 3}+{\pi^2 \over 18}{m_{\tilde g}^2\over 2 k_1\cdot k_2}
-{1\over 3}{m^2_{\tilde g}\over 2 k_1 \cdot k_2}
\textrm{Li}_2\(1-{2 k_1\cdot k_2 \over m^2_{\tilde g}}\)\right.\nonumber \\
& & \left. 
~~~~~~~~~~~~~~~~~~~+{1\over 6}\(1+{m_{\tilde g}^2\over 
m_{\tilde g}^2-2 k_1\cdot k_2}\)
\log \({2 k_1\cdot k_2 \over m^2_{\tilde g}}\)\]. 
\eea
Due to the chiral structure of the loop, the result is
proportional to the mass of the internal fermions. 

Using these expressions and the definition of $g_{eff}^{L/R}$ 
in~(\ref{eq:sqqa}), we arrive at
\bea
\label{eq:gefffull}
g_{eff}^{L/R}
&=&
\mp i {4\sqrt{2} \over 3} {\alpha_s^2 \over \fa} 
\left(I_1^{L/R} - 2 I_2^{L/R} \right) \\
&\simeq&
\mp {\alpha_s^2 \over \sqrt{2} \pi^2} 
{m_{\tilde g}\over \fa} \log\({\fa\over m_{\tilde g}}\),
\label{eq:geffapprox}
\eea

Note that the dominant contribution comes from the 
logarithmically divergent part. However, as in all supersymmetric
theories the dependence on the cut-off is weak. Also, the
limit of large $\fa $ is perfectly well-defined and the coupling
vanishes in the decoupling limit, when $\fa \rightarrow +\infty$.
Another important point is that the coupling depends on a
single unknown parameter, the ratio $m_{\tilde g}/ \fa$: 
therefore in the following we keep $\fa = 10^{11} \gev $
fixed, and vary the value of the gluino mass to 
explore the parameter space.

Secondly, the effective vertex is proportional to
the gluino mass and tends to zero in the limit of exact supersymmetry and
massless quarks; this can be easily understood by considering
that for massless quarks the whole axion multiplet decouples from
the low energy Lagrangian~\cite{axionreviews:cite}. 
When the chiral symmetry is broken, a term proportional to the 
quark mass survives, giving an axino-quark-squark coupling of
the same type as the standard axion-quark-quark coupling.

However, in the case of broken supersymmetry a new contribution
to the coupling arises from the gluino mass and dominates
for all light quarks. For the top quark,
the term proportional to $m_{t}$ could be important for
small gluino masses. We neglect this term here since it would not 
change our results substantially.

It is also worth mentioning that the full expression above contains, 
in the limit of negligible final state masses, terms proportional
to $\log(1-m_{\tilde q}^2/m_{\tilde g}^2)$. 
Therefore $g_{eff}^{L/R}$ contains an imaginary component 
if $m_{\tilde q} > m_{\tilde g}$. This imaginary piece
corresponds to the ``cut" loop diagram, in which the intermediate 
gluino and quark are \emph{real}. Since this contribution is already 
included when considering the squark decay into a gluino and a quark, 
we keep only the real part of $g_{eff}^{L/R}$ to avoid double counting.

%%%%%%%%%%%%%%%%%%%%%%%%%%%%%%%%%%%%%%%%%%%%%%%%%%%%%%%%%%%%%%%
\section{Thermal production of axinos}\label{sec:thermalaxino}

For reheat temperatures below $\fa $, the axinos are not in 
thermal equilibrium with the MSSM particles. Assuming that any 
primordial axino population was completely diluted away during an 
inflationary stage, the present axino density has been generated
after the end of inflation, since the epoch of reheating/preheating.
Two different production mechanisms are possible: {\it thermal production}, 
TP, in which axinos are produced via thermal processes in the primordial 
plasma; and {\it non-thermal production}, NTP, which is independent 
of the presence of the thermal bath.
An example of the latter is the decay of the neutralino 
NLSP after it freezes out, as considered in~\cite{ckr}.
In this case the final number density is related only to the
density and lifetime of the initial neutralino population.

In the case of TP, on the other hand, the axino number density 
$\naxino$ depends on the species and processes present in
the plasma, and can be obtained by integrating the Boltzmann 
equation with the appropriate collision integral. 
Fortunately,  since the number density of axinos is well below 
the equilibrium one for $\treh \ll \fa$, we can neglect
the inverse processes, which are suppressed by $\naxino$, and we have:
\begin{equation}
\label{eq:Boltzmann}
\frac{d\,\naxino}{d\,t} + 3H \naxino =
\sum_i \langle\Gamma(i\ra \axino+\cdots)\rangle n_i +
\sum_{i,j} \langle\sigma(i+j\ra \axino+\cdots)\vrel\rangle n_i n_j,
\end{equation}
where the Hubble parameter, $H$, can be given as a function of the
temperature by 
$H(T)=\left[\left(\pi^2\gstar\right)
/\left(90\mplanck^2\right)\right]^{1/2}\;T^2$  
in the radiation-dominated era, 
with $g_{*}$ counting the number of effective massless degrees of freedom
and $\mplanck$ denoting the reduced Planck mass, 
$\mplanck = 2.4 \times 10^{18} \gev $.

The first term on the r.h.s. takes into account all the decay channels 
of the  $i$th particle with one axino in the final state, and
$\langle\Gamma(i\ra \axino+\cdots)\rangle $ is the thermally averaged 
decay width for the process.
The second term corresponds to axino production via 2 to 2
particle scatterings, with $\sigma(i+j\ra \axino+\cdots)$
denoting the cross-section for particles $i,j$ scattering 
into final states involving axinos and $\vrel$ the
relative velocity.
Averaging over initial spins and summing over final spins is understood.

In order to solve Eq.~(\ref{eq:Boltzmann}), it is convenient to 
separate the two contributions and introduce 
the axino TP yield as~\cite{ckkr} 
\begin{eqnarray}
\ythaxino
&=& \frac{\naxino^{\rm TP}}{s} \nonumber \\
&=& \sum_{i}\yadec + \sum_{i,j}\yascat\, ,
\label{ythaxino:eq}
\end{eqnarray}
where $s= (2\pi^2/45)\gsstar T^3$ is the entropy density, 
and normally $\gsstar=\gstar$ in the early Universe. 

Moreover, by changing variables from the cosmic time $t$ to the temperature 
$T$,
we can write the two solutions of the Boltzmann differential equation 
easily in integral form:
\bea
\label{eq:Ydec}
\yadec (T_{0}) &=& 
\int_{T_{0}}^{\treh}dT\,
\frac{\langle\Gamma(i\rightarrow\axino +\cdots)\rangle n_i}{sHT}\\
\label{eq:Yscat1}
\yascat (T_{0}) &=&
\int_{T_{0}}^{\treh}dT\,
\frac{\langle\sigma(i+j\rightarrow\axino +\cdots)\rangle n_in_j}{sHT}.
\eea
where we have considered the evolution from the reheating temperature 
after inflation, $\treh$, down to the present temperature $T_{0}$.
Full expressions for $\yadec $ and $\yascat$ can be found in~\cite{chkl}.

In the following we consider the axino yield generated by the 
effective coupling in Eq.~(\ref{eq:sqqa}). We compute the contribution due
to squark decays and estimate the additional contribution 
to quark scattering.
The axino number density arising from gaugino decays 
and scattering via the dimension-$5$ gluino-gluon-axino 
operator in~(\ref{eq:agg}) was computed in~\cite{ckkr}.

%%%%%%%%%%%%%%%%%%%%%%%%%%%%%%%%%%%%%%%%%%%%%%%%%%%%%%%%%%
\section{Squark decay}\label{sec:sqdecay}

The squark can decay into axinos in two different ways: via
$g_{eff}^{L/R}$ into a two-body final state of an axino and a quark,
or via an exchange of an intermediate gluino/neutralino and the
anomalous couplings into a three-body final state of an axino, a quark
and a gluon/photon. As we will see, the first one will play an
important role while the second will always remain subdominant.

%+++++++++++++++++++++++++++++++++++++++++++++++++++++++++
\subsection{Two-body squark decay}\label{sec:twobodydecay}

The process involves the diagrams shown in Fig.~\ref{fig:sqqa}. 
The width for this decay of both left- and right-handed squarks
is given by 
\bea
\label{eq:gammasquark}
\Gamma_{\tilde q}^{2body}
&=& 
{(g_{eff})^2\over 16\pi} m_{\tilde q} \nonumber\\
&=& 
{\alpha_s^4 \over (2\pi)^5 } m_{\tilde q} 
\[{m_{\tilde g}\over \fa} \log \({\fa\over m_{\tilde g}}\)\]^2 
\label{eq:decay-2b}\\
&\simeq& 
 \nonumber
\left(2.6 \times 10^{-6}\,\textrm{sec}\right)^{-1}
\({m_{\tilde q}\over 500 \gev}\)
\[\({m_{\tilde g}\over 1 \tev}\)
\(1- 0.05\log \({m_{\tilde g}\over{{1\tev}}}\)\)\]^2\! ,
\eea
where in the last line we have taken the default value of $\fa =
10^{11}\gev$.  This process is suppressed by a large power of the
strong coupling and by loop factors, but it can still be sizable for
large gluino mass, or equivalently for smaller $\fa$.

%+++++++++++++++++++++++++++++++++++++++++++++++++++++++++++
\subsection{Three-body decay}\label{sec:threebodydecay}

\begin{figure}[h]
  \begin{center}
    \epsfig{file=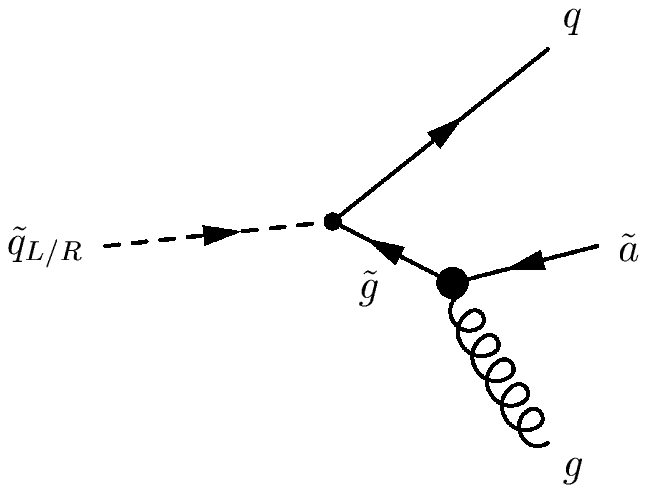, height=2.3in}      
    \label{fig:loop2}
    \caption{Three-body decay of a squark into an axino, a quark and a
    gluon. The thick dot denotes the effective gluino-gluon-axino
    vertex of Fig.~1. %\protect{\ref{fig:loop0}}. 
	}
  \end{center}
\end{figure}

At an effective tree-level, the squark will 
also decay into an axino, a quark
and a gluon, as illustrated in Fig.~\ref{fig:loop2}.

%%%%%%%%%%%%%%%%%%%%%%%%%%%%%%%%%
%%%%%%%%%%%%%%%%%%%%%%%%%%%%%%%%

The decay rate for this process in the case
of large gluino mass $m_{\tilde g} > m_{\tilde q} $
and negligible final state masses is:
\beq
\Gamma_{\tilde q}^{3body} = {4 \alpha_s^3 \over 3 (4 \pi)^4 }  
{m^3_{\tilde q}\over \fa^2}
F \({m_{\tilde g}^2 \over m_{\tilde q}^2} \)
\label{eq:decay-3b}
\eeq
where we have summed over the color indices of the final state gluon
and quark and where
\beq
F (x) = {1\over 6} -{13 x\over 4} +{7 x^2\over 2 }+
{x\over 4}  \(1 - x \) \(3 - 7 x \) \log\[{(1-x)^2\over x^2}\].
\eeq
In the limit $m_{\tilde g} \gg m_{\tilde q} $, (\ref{eq:decay-3b}) reduces to
\beq
\Gamma_{\tilde q}^{3body} = {\alpha_s^3 \over 18 (4\pi)^4 }  
{m^5_{\tilde q}\over m^2_{\tilde g}\fa^2}
~\[1 + O\({m_{\tilde q}^2\over m_{\tilde g}^2} \) \] .
\label{eq:decay-3blimit}
\eeq
Note again that, in our effective theory, 
the 3-body decay width~(\ref{eq:decay-3b}) is
proportional to the cube of the mass of the decaying particle. 
However, in the limit when one can neglect
momentum flow in the intermediate states, one recovers
in~(\ref{eq:decay-3blimit}) the usual fifth 
power of mass for a dimension-4 interaction.  

Comparing~(\ref{eq:decay-3blimit}) with~(\ref{eq:decay-2b}) gives  
\beq
{\Gamma_{\tilde q}^{3body} \over \Gamma_{\tilde q}^{2body}}
= {\pi \over 144 \alpha_s } \;
\left({m_{\tilde q}\over m_{\tilde g}}\right)^4\; 
\log^{-2} \(\fa/m_{\tilde g}\) 
\ll 1
\eeq
The $3$-body decay channel is therefore subdominant for squarks
lighter than gluinos. 
For example, if $m_{\tilde q} = 500 \gev, m_{\tilde g} = 1 \tev$
and $\fa = 10^{11} \gev $, we have
\bea
\Gamma_{\tilde q}^{3body}
&=& \left(4.1 \times 10^{-2}\,\textrm{sec}\right)^{-1}\, ,  
\eea
which is suppressed by about four orders of magnitude relative to 
the two-body decay~(\ref{eq:decay-2b}).

An analogous decay channel exists with an intermediate neutralino
replacing the gluino
and a final photon rather than a gluon. 
Taking the neutralino to be a bino~\cite{binodm}, the decay of each color 
of squark into a quark, an axino and a photon is given by:
\beq
\Gamma_{\tilde q}^{L/R}= 
{\alpha_{em}^3 \over (4\pi)^4} 
~\({\cayy Z^2_{11} Y_{\squark^{L/R}} \over \cos\theta_W}\)^2 
~\({m^3_{\tilde q}\over \fa^2}\)
~F \({m_{\neut}^2\over m_{\tilde q}^2} \),
\eeq
where
$Y^2_{\squark^{L/R}} = 1/36, 4/9, 1/9 $ for a left-handed squark,
a right-handed up-type squark and a right-handed down-type squark,
respectively.
For $m_{\tilde q} = 500 \gev, m_{\tilde g} = 1 \tev$
and $\fa = 10^{11} \gev $, we have
\beq
\Gamma_{\tilde q}^{L/R} \simeq
\left(10\, \textrm{sec}\right)^{-1} \;
\cayy^2 Z^2_{11}  Y^2_{\squark^{L/R}}
\eeq
This is an even smaller contribution than that from the gluino
exchange diagram above. Therefore we conclude that  
three-body decays are not expected to contribute 
substantially to the axino yield.

Note that we have considered the decays only for the case of
intermediate particle heavier than the initial squark.
For an intermediate particle lighter than the initial squark,
the dominant contribution 
to the axino production comes from the resonance.
This contribution can be factorized into the
squark decay to a \emph{real} neutralino/gluino, and its 
subsequent decay to an axino and a photon/gluon.
This process is already included in the Boltzmann equation in the 
gaugino/neutralino decay term.

%++++++++++++++++++++++++++++++++++++++++++++++++++++++++++++++++
\subsection{Yield from the decay of squarks}\label{sec:threeodyyield}

In general, the thermal average of a decay described by a width $\Gamma$ is
given by
\begin{eqnarray}
\langle\Gamma\rangle n_i =  
\Gamma~\frac{m_i T^2}{2 \pi^2} 
\int_{m_i/T}^\infty dx \, \frac{(x^2-m_i^2/T^2)^{1/2}}{e^x \mp 1}
\label{eq:tagam}
\end{eqnarray}
where the $-(+)$ is for decaying boson (fermion). 
By inserting this expression into~(\ref{eq:Ydec}), one obtains
\begin{eqnarray}
\label{eq:Ydec2}
\yadec & = &\frac{\gbar M_P \Gamma}{16 \pi^2 m_i^2}
\int_{t_R}^{\infty}dt\,\frac{1}{e^t \mp 1}
\Bigg[\Bigg\{\frac{\pi}{2}-\textrm{tan}^{-1}\Bigg(
\frac{t_R}{\sqrt{t^2-t_R^2}}
\Bigg)\Bigg\}t^4 \nonumber \\
&  & \hspace{0.8in}
+ t_R (t^2-2t_R^2)\sqrt{t^2-t_R^2}~\Bigg]
\end{eqnarray}
where $t_R=m_i/T_R$ and $\gbar=135\sqrt{10}/(2\pi^3 g_{*}^{3/2})$,
for $g_{*}=915/4$ for the MSSM in the early Universe. 

In the case
$T_R > m_i$, the integration gives the simple result
\begin{equation}
\label{eq:Ydec3}
\yadec \simeq \frac{3 \zeta(5) \gbar M_P \Gamma}{4 \pi m_i^2}.
\end{equation}

To obtain the axino yield from squark decay, we 
substitute~(\ref{eq:gammasquark}) for the decay rate above, and obtain 
the following expression, valid for $T_R > m_{\tilde q}$
\bea
Y_{\tilde q}^{dec} 
&=& 
\frac{6 \zeta (5) \alpha_s^4}{(2 \pi)^6}\,\gbar\,\(\frac{M_P}{m_{\tilde q}}\) 
\[\( \frac{m_{\tilde g}}{\fa}\) \log\(\frac{\fa}{m_{\tilde g}}\)\]^2 \\
&\simeq& 
3~\(\frac{m_{\tilde g}}{m_{\tilde q}}\) Y_{\tilde g}^{dec} \simeq 10^{-8}.
\eea
The numerical value corresponds to $m_{\tilde g} = m_{\tilde q} = 1 \tev$
and $\fa = 10^{11}\gev $. 
It is clear that squark decay is typically as important 
as gluino decay in the thermal production of axinos, and can be substantially
more important if the squarks are lighter than the gluinos.
For $T_R$ less than the mass of the decaying sparticle, both yields are 
Boltzmann suppressed and negligible.

%%%%%%%%%%%%%%%%%%%%%%%%%%%%%%%%%%%%%%%%%%%%%%%%%%%%%%%%%%
\section{Scattering of quarks and squarks}\label{sec:dim4}

The effective dimension-$4$ squark-quark-axino
coupling~(\ref{eq:gefffull}) also gives new contributions to some of
the scattering axino production 
processes previously considered in~\cite{ckkr} which
involved the effective gluino-gluon-axino vertex~(\ref{eq:agg}).
However, we shall see that such additional contributions will be 
always subdominant.

%%%%%%%%%%%%%%%%%%%%%%%%%%%%%%%%%
\begin{figure}[t]
  \begin{center}
    \epsfig{file=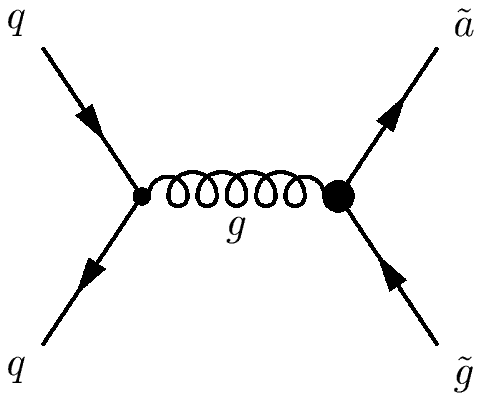, height=1.5in}      
    \caption{Example of dimension-5 contribution 
		to the process $qq\rightarrow \squark\axino$ (Process~I).
            }
    \label{fig:loop4}
  \end{center}
\end{figure}
%%%%%%%%%%%%%%%%%%%%%%%%%%%%%%%%

As an example and for a conservative estimate, let us consider 
the scattering of two quarks into a 
gluino and axino. (This process was called process~I in~\cite{ckkr}.) 
This process does not contain sparticles in the initial state and is
therefore never Boltzmann-suppressed by the number density of the
initial particles.\footnote{Nevertheless, there is some suppression
coming from final state gluino mass.} Therefore process~I is expected 
to be one of the dominant contributions to the axino yield from
dimension-4 interactions at low temperatures.               

As we will show below, the dimension-$5$ diagrams that contribute to the 
scattering process, illustrated in Fig.~\ref{fig:loop4},
are of order $g_s^3$~\cite{ckkr}. The two dimension-$4$ diagrams
that contribute are of order $g_s^5$. They are shown in
Fig.~\ref{fig:loop5}, where the shaded squares denote the effective
1-loop 
squark-quark-axino vertices given in Fig.~\ref{fig:sqqa}. 
Although these two diagrams are
suppressed by $g_s^2$ relative to the dimension-$5$ contributions, the
enhancement through the large logarithm and the gluino mass present in
$g_{eff}^{L/R}$ partially compensates for it.

Generalizing the cross section due to the dimension-$5$
coupling computed in Ref.~\cite{ckkr} by including a full dependence
on the gluino mass, we find that
\beq
\sigma_{I,5} =
{\alpha_s^3 \over 72 \pi^2 \fa^2}\,\[1-3\({m_{\tilde g}^2 \over s}\)^2
+2\({m_{\tilde g}^2 \over s}\)^3\] ,
\eeq
where the eight gluino species in the final state have been summed
over, and the six flavors of quark in the initial state have been
averaged over.

The combined contribution from the two dimension-$4$ channels is
\bea
\sigma_{I,4}
=
\frac{2 \alpha_s (g_{eff}^{L/R})^2}{3 s^2}
\[\frac{(s-m_{\tilde g}^2)\[s-2(m_{\tilde g}^2-m_{\tilde q}^2)\]}
{s+m_{\tilde q}^2-m_{\tilde g}^2}
-(2 m_{\tilde q}^2-m_{\tilde g}^2)
\log\(\frac{s+m_{\tilde q}^2-m_{\tilde g}^2}{m_{\tilde q}^2}\)\]
\label{sigma-dim4}
\eea
and is proportional to $\alpha_s^5$. 
The two squarks corresponding to each quark and the eight gluino
species have been summed over in this expression, while we
averaged over the initial flavor.

%%%%%%%%%%%%%%%%%%%%%%%%%%%%%%%%%
\EPSFIGURE{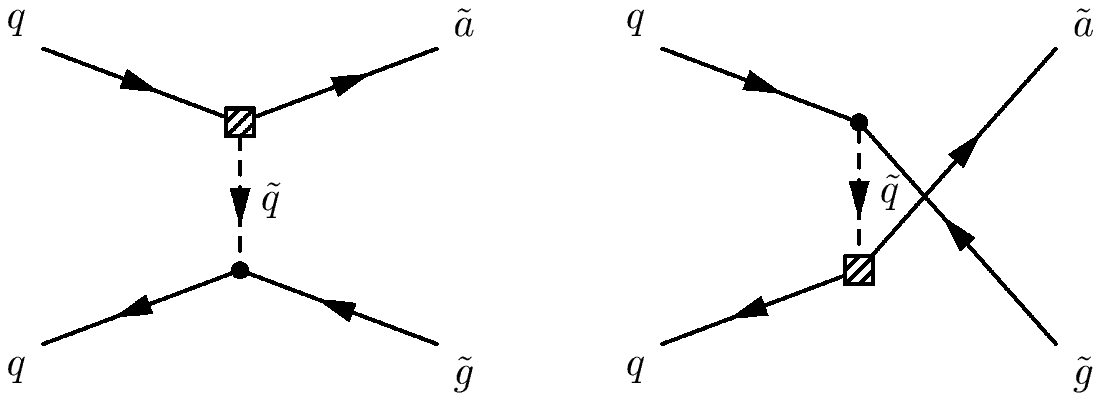, height=1.6in}      
{Dimension-4 contributions to Process~I.
\label{fig:loop5}
}
%%%%%%%%%%%%%%%%%%%%%%%%%%%%%%%%

Interference between the dimension-$4$ $t$ and $u$ channels gives a
contribution proportional to $\maxino$, which is neglected here.

There is also a contribution to the cross-section from
interference between dimension-$4$ and dimension-$5$ diagrams, which
is of order $\alpha_s^4$.  Giving the $L$ squarks (${\widetilde q}_L$)
a common mass $m_L$ and the $R$ squarks (${\widetilde q}_R$) 
a common mass $m_R$, this 4/5 interference term can be written as 
\bea
\sigma_{I,4/5}%& = &
=\frac{\sqrt{2}\,\alpha_s^2 \,|g^{L/R}_{eff}|}{3 \pi \fa}
\(\frac{m_{\tilde g}}{s^2}\) 
\[m_L^2\,\log\(\frac{s+m_L^2-m_{\tilde g}^2}{m_L^2}\) 
-m_R^2\,\log\(\frac{s+m_R^2-m_{\tilde g}^2}{m_R^2}\)\]
\label{eqn:45int}
\eea
Note that the contributions from the $L$ and $R$ squarks 
cancel in the limit $m_L=m_R$ thanks to the opposite sign of the
coupling $g^{L/R}_{eff}$. 

%++++++++++++++++++++++++++++++++++++++++++++++++++++++++++
\subsection{Yield from dimension-$4$ scattering processes}

We numerically computed the axino yield using an estimate of
dimension-4 contribution to Process I derived above, and we found 
it to be negligible compared to the dimension-5 part for any reheat
temperature. 

To illustrate this analytically, let us consider the following
simplified example.  In the limit $m_{\tilde q}=m_{\tilde g}=m$, the
contribution to the Process~I cross-section from the dimension-$4$
channel is
\beq
\sigma_{I,4}=\frac{2 \alpha_s (g_{eff}^{L/R})^2}{3 s^2}
\[s-m^2-m^2\log\(\frac{s}{m^2}\)\]
\eeq
To calculate the yield we expand around $m/T_R =0$.  After summing
over the four possible combinations of the initial spins, the six
quark species and the three colors, and multiplying by a number
density factor $3/4$ for each initial fermion, we find
\bea
Y_{I, 4}^{scat} &=& \frac{9\alpha_s (g_{eff}^{L/R})^2~\gbar}{8\pi^3}
~\frac{M_P}{m} +O\(\frac{m}{T_R}\)\\
&\simeq& 4.9 \times 10^{-10} \(\frac{\alpha_s(m)}{0.11}\)^5
\(\frac{10^{11}\gev}{\fa}\)^2 \(\frac{m}{1\tev}\).\nonumber
\label{yscatt-4}
\eea
So for $T_R$ greater than the masses of the external particles, 
the axino yields from these dimension-$4$ processes are 
approximately independent of $T_R$. 
For this reason, in analogy with decay processes, we have used 
$\alpha_s(T=m)$ to take account of the running of the strong coupling.

This is in contrast with what happens for 
the dimension-$5$ scattering processes. For example,
the contribution of the diagram in Fig.~\ref{fig:loop4} to the yield is
\bea
Y_{I, 5}^{scat} &\simeq& \frac{9 \alpha_s^3~\gbar M_P}{16 \pi^6 \fa^2} T_R \\
&\simeq& 3.7 \times 10^{-10}\,\(\frac{\alpha_s(T_R)}{0.11}\)^3
\(\frac{10^{11}\gev}{\fa}\)^2\,\(\frac{T_R}{1\tev}\) \nonumber
\eea
for high reheat temperatures, i.e. $Y_{I,5}\propto T_R$.  This different 
behavior is simply due to dimensional reasons: the
dimension-$5$ scatterings have cross-sections that are proportional to
$1/\fa^2$ and independent of $s$ at large $s$, while the dimension-$4$
scatterings are characterized by a dimensionless coupling
$g_{eff}^{L/R} \propto m_{\tilde g}/\fa $, and their cross-sections are
suppressed by $s^{-1}$ at high energies. 

Comparing the two yields, we find that for $T_R\gg m$
\beq
Y_{I, 4}^{scat} \simeq 1.3 \,Y_{I, 5}^{scat}
\,\(\frac{\alpha_s(m)}{0.11}\)^5 
\(\frac{0.11}{\alpha_s(T_R)}\)^3 \(\frac{m}{T_R}\)
\eeq
Therefore the contribution to the yield from the two dimension-$4$
diagrams is considerably less than the contribution from the
dimension-$5$ diagram already considered for large reheat
temperature. As we will see, this is the case also for $T_R \simeq m$,
when the analytic formulas above are no more valid, since the
dimension-$4$ yield reaches the asymptotic value in
eq.~(\ref{yscatt-4}) more slowly than the dimension-$5$ one.
 
Below eq.~(\ref{eqn:45int}) we noted that the contributions to the 
cross-section from interference between dimension-$4$ 
and dimension-$5$ diagrams cancel in the limit $m_L=m_R$.
However, in the case of unequal masses there is a remainder, and therefore 
a contribution to the yield from $4/5$ interference.  
We have calculated this contribution in two extreme cases;
\begin{eqnarray}
Y^{scat}_{I,4/5} &\simeq&  ~~\frac{2}{3}\, Y^{scat}_{I,4}~~~\textrm{if}~~
m_L=10\, m_R \\
& & \nonumber \\
&\simeq&  -\frac{2}{3}\, Y^{scat}_{I,4}~~~\textrm{if}~~m_L=0.1\, m_R.
\end{eqnarray}
In either case, the magnitude of the effect of the dimension-$4$ diagrams does
not become comparable with that of the dimension-$5$ diagrams.

%%%%%%%%%%%%%%%%%%%%%%%%%%%%%%%%%
\begin{figure}[htp]
  \begin{center}
    \epsfig{file=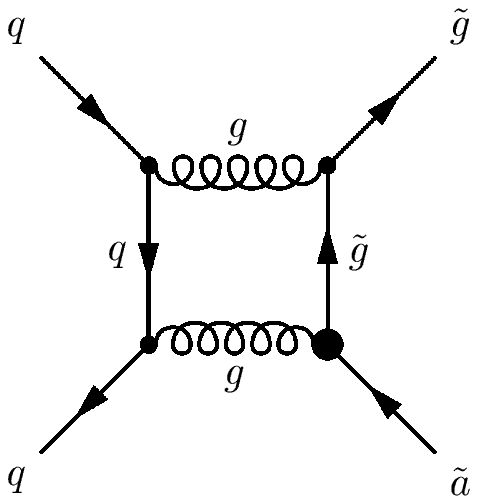, height=1.5in}      
    \caption{An example of a 2-loop contribution to Process~I.}
    \label{fig:loop5a}
  \end{center}
\end{figure}
%%%%%%%%%%%%%%%%%%%%%%%%%%%%%%%

In principle, there are several other diagrams of order $\alpha_s^5$ that 
should be computed. 
For example, the box diagram in Fig.~\ref{fig:loop5a} will contribute to the 
total cross-section of Process~I. 
Such diagrams are also 2-loop in a renormalizable theory (and
therefore 1-loop in the effective theory) 
and as a result suppressed by the same loop 
factors, but contain no logarithmic divergence and so do not receive 
the enhancement factor contained in the diagrams considered above.
Therefore it seems reasonable to neglect them too.

We conclude that dimension-$4$ scattering diagrams never give a significant 
contribution to the thermal yield of axinos.
The dominant contributions 
to $Y_{\tilde a}^{scat}$ are always those from dimension-$5$ diagrams
already considered in~\cite{ckkr}.

%%%%%%%%%%%%%%%%%%%%%%%%%%%%%%%%%
\EPSFIGURE{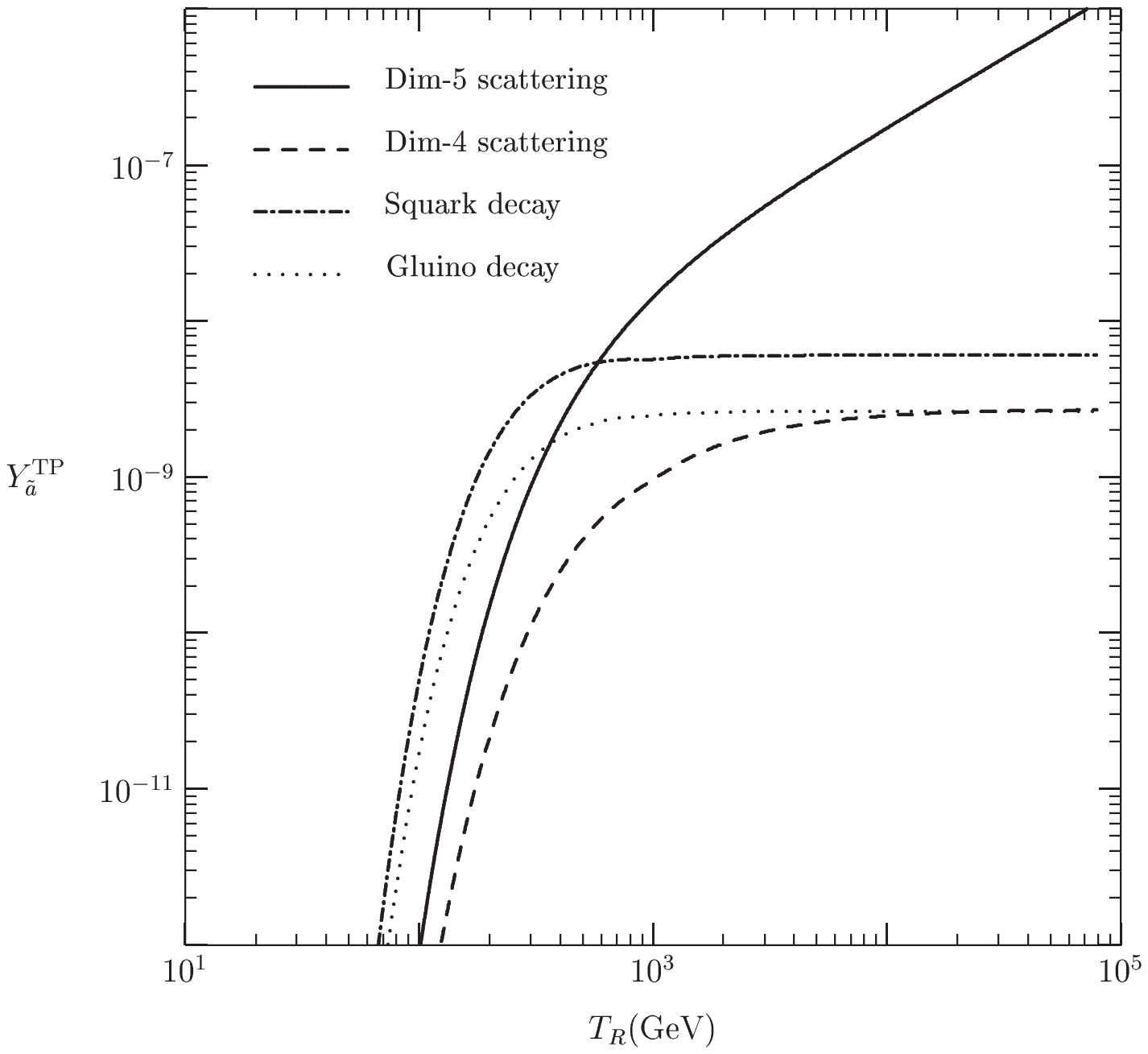, height=4.0in}      
{The yield of axinos from thermal production, $Y_{\tilde a}^{TP}$,
     as a function of the reheat temperature $T_R$, assuming
     $\fa=10^{11}\gev$ and $m_{\tilde q}=m_{\tilde g}=1\tev$. 
    \label{fig:ytr}
}
%%%%%%%%%%%%%%%%%%%%%%%%%%%%%%%% 

%%%%%%%%%%%%%%%%%%%%%%%%%%%%%%%%%%%%%%%%%%%%%%%%%%%%%%%%%%%%
\section{Results}\label{sec:results}

The relative contributions from the processes considered above 
to the axino abundance are summarized in
Fig.~\ref{fig:ytr}.  The solid line is the yield from dimension-5
scattering processes, while the dashed line is an estimate of
dimension-4 scattering processes, computed by multiplying $Y_{I,
4}^{scat}$ by seven, which is the number of channels receiving
dimension-4 contributions.  The figure illustrates that, for equal
squark and gluino masses, the yield from dimension-4 scattering
processes is negligible for any value of $T_R$, whereas the yield from
the decay process $\squark \to q \axino$ can dominate the abundance
for $T_R \lsim m_{\tilde q}$.

%%%%%%%%%%%%%%%%%%%%%%%%%%%%%%%%%%%%%%%%%%%%%%%%%%%%%%%%%%%%%%%
\EPSFIGURE{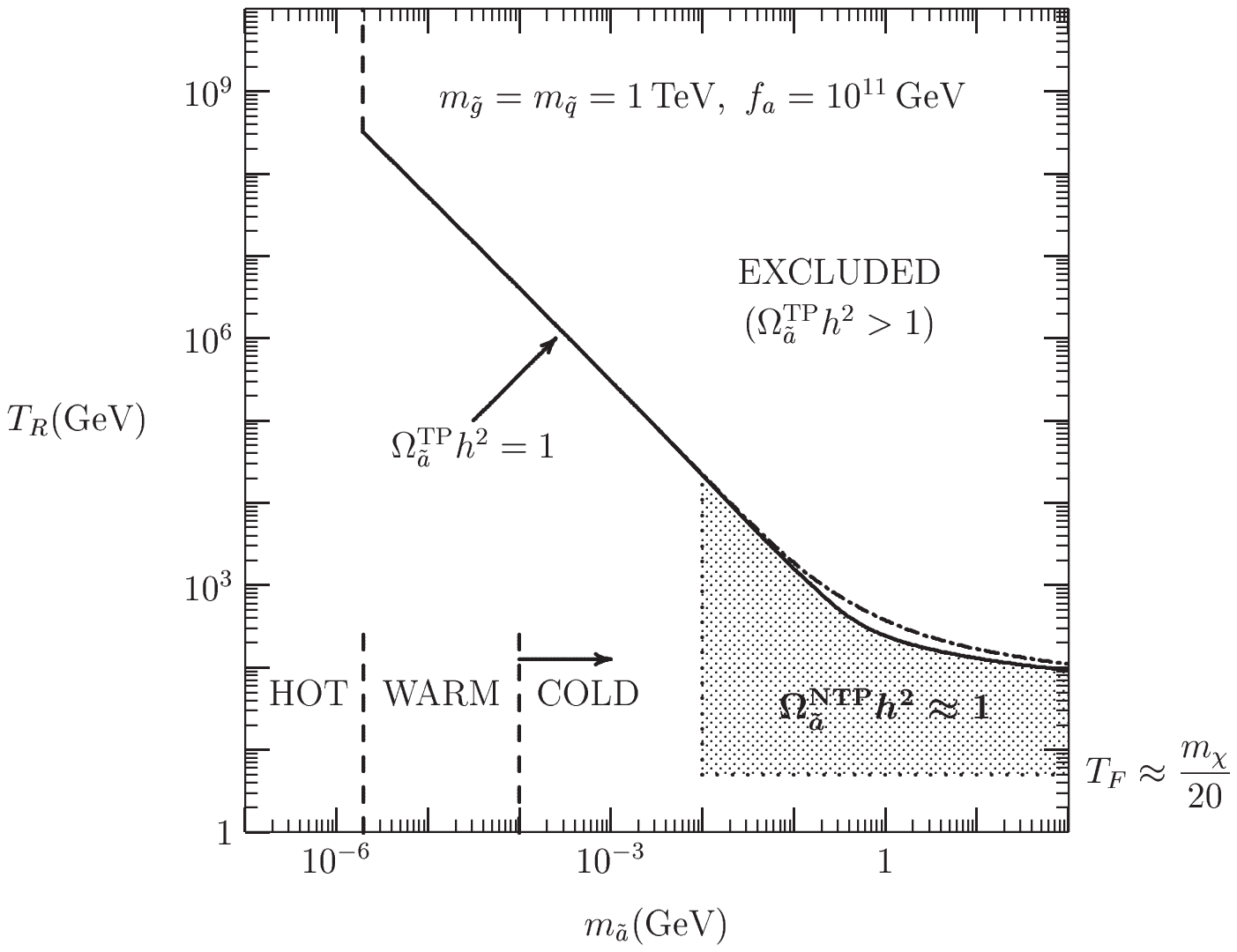, width=5.0in}     
{The ($\maxino, \treh$) plane, with a bino-like neutralino of
         mass $100\gev$, $m_{\tilde g}=m_{\tilde q}=1\tev$ and $F=10^{11}\gev$.
         The dash-dotted line represents the bound if squark decay is 
         neglected, while the thick line is the full result.  
\label{fig:trmax} 
}
%%%%%%%%%%%%%%%%%%%%%%%%%%%%%%%%%%%%%%%%%%%%%%%%%%%%%%%%%%%%%%

The present fraction of the axino energy 
density to the critical density $\Omega_{\tilde a}$ is given by
\begin{equation} 
\maxino \yaxino \simeq 0.72 \ev \(\abunda \over 0.2\)
\label{eq:m-yaxino}
\end{equation}
where $h$ is the Hubble
parameter. Using this relationship, one can redisplay the
results of Fig.~\ref{fig:ytr} in the ($\maxino, \treh$) plane. 

%%%%%%%%%%%%%%%%%%%%%%%%%%%%%%%%%%%%%%%%%%%%%%%%%%%%%%%%%%%%%%%
\EPSFIGURE{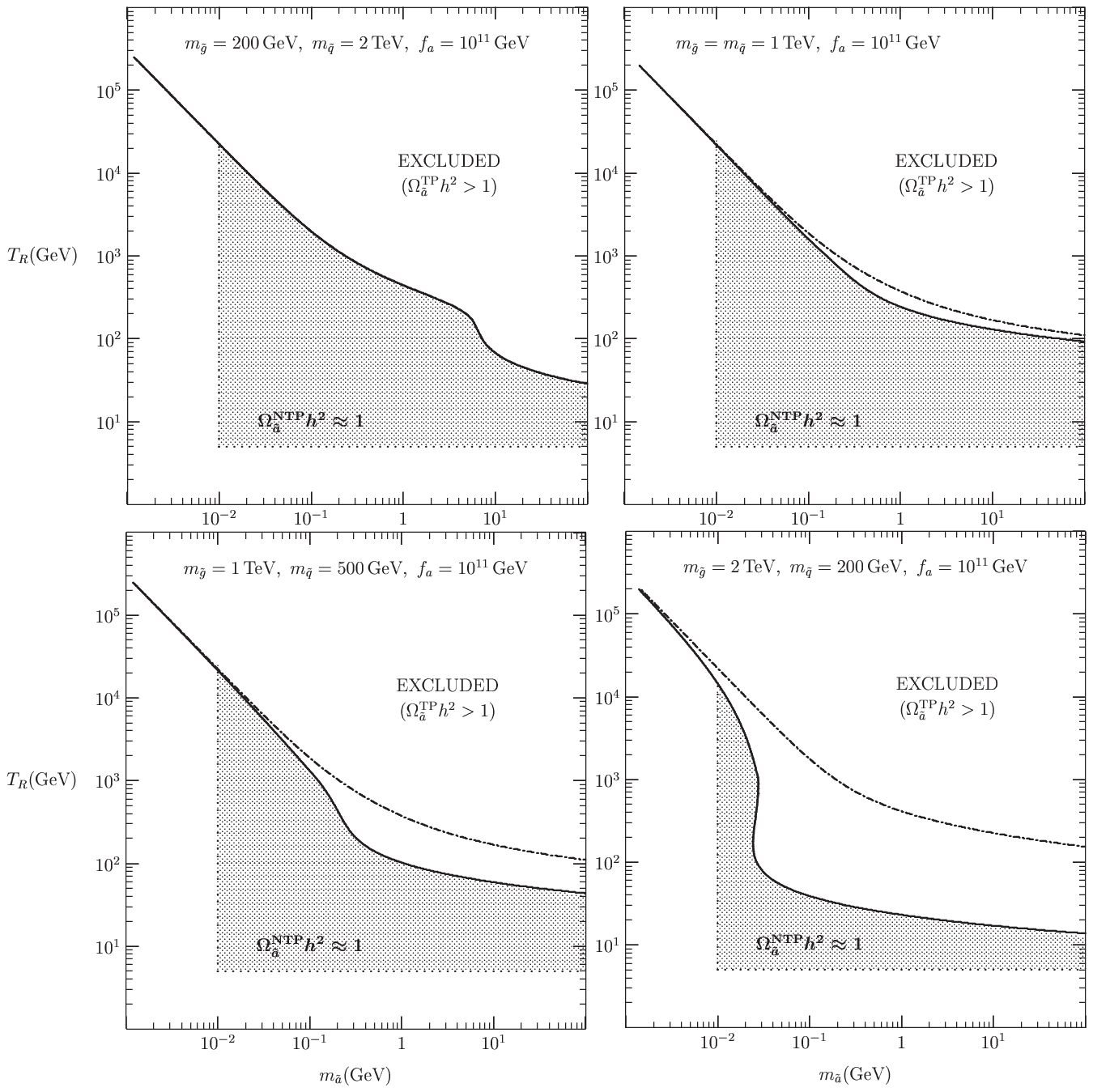,height=6in, width=\textwidth}
{The ($\maxino, \treh$) plane, showing the different bounds
on the reheat temperature depending on the mass spectrum.
For light gluinos, the effective coupling is small and
its effect on the axino abundance negligible. 
For light squarks, axino production through squark 
decay is so large that most of the allowed parameter space is
excluded.
Note that, in the figure in the left top panel, the kink at
$\treh \simeq 10^2 \gev$ is where gluino decays 
start to dominate scatterings. For large gluino masses
the kink is moved to the left and is less pronounced.
In the right bottom panel, a similar kink is due instead
to the decay of light squarks.
\label{fig:trmax2}
} 
%%%%%%%%%%%%%%%%%%%%%%%%%%%%%%%%%%%%%%%%%%%%%%%%%%%%%%%%%%%%%%%%%%%%%%%%

For axinos produced through scatterings and decays of particles in the
plasma (TP axinos), eq.~(\ref{eq:m-yaxino}) gives an upper bound on
$\treh$ as a function of the axino mass from the requirement
$\omegaatp h^2 \lsim 1$. This bound is depicted by a thick solid 
line in Fig.~\ref{fig:trmax},
for $m_{\tilde g}=m_{\tilde q}=1\tev$ and $F=10^{11}\gev$.

At large $\treh$, the TP mechanism is dominant and the
cosmologically favored region ($0.2\lsim\abunda\lsim0.4$) forms a
narrow strip (not indicated in Fig.~\ref{fig:trmax})
just below the $\omegaatp=1$ boundary. 

For reheat temperatures less than the mass of the gluinos and squark, 
non-thermal production of axinos via out of equilibrium 
neutralino decays can dominate~\cite{ckr, ckkr}. In that case the neutralinos 
decay into an axino and a photon or Z well after freeze-out, and the final 
axino abundance is independent of the reheat temperature, as 
long as this is high enough for the neutralinos to thermalize before 
freeze-out.
The shaded region below the bound on $\treh $, shows where
a sufficiently large axino population can be obtained
by the decay of a nearly pure bino  with mass $\mchi=100\gev$.
The ranges of $\maxino$ where axinos are hot/warm/cold 
dark matter are also shown. More details can be found in~\cite{ckkr}.

In Fig.~\ref{fig:trmax2}, an enlarged region of the ($\maxino, \treh$) plane
is shown for different values of the sparticle masses. The dash-dotted 
line always represents the bound if squark decay is neglected;
clearly, for heavy gluinos, a large region
of the parameter space is excluded by including squark decay.
This remarkably large effect is due to two factors: for one thing, a light 
squark remains in thermal equilibrium and can play a role in axino production 
down to lower temperatures; for another, a large gluino mass enhances the 
squark decay to axinos.

Note that, as discussed previously, the squark decay
rate into axinos is only regulated by the ratio $m_{\tilde g}/\fa $.
Therefore the same effect could be due to a smaller effective 
PQ scale $\fa $ rather than a larger gluino mass.

%%%%%%%%%%%%%%%%%%%%%%%%%%%%%%%%%%%%%%%%%%%%%%%%%%%%%%%%%%%%
\section{Conclusions}\label{sec:concl}

We have computed the effective axino-quark-squark vertex
arising in KSVZ axion models through one loop diagrams,
and we considered the effects of this interaction 
for the  thermal production of axinos.

As expected by dimensional arguments, we find that the
dimension-$4$ scattering diagrams generated
by this coupling are negligible with respect to the
dimension-$5$ scattering processes already analysed
in~\cite{ckkr}. On the other hand, the two body decay of a squark 
to an axino and a quark can be the dominant axino
production channel at low reheat temperature, in
particular when the squarks are substantially lighter
than the gluinos. Our results put tighter constraints
on the reheat temperature for the axino CDM scenario,  
summarized in Fig.~\ref{fig:trmax2}. For very light
squark masses and heavy gluinos, most of the parameter
space for NTP axino DM is excluded.

\clearpage

%%%%%%%%%%%%%%%%%%%%%%%%%%%%%%%%%%%%%%%%%%%%%%%%%%%%%%%%%%%%
\acknowledgments

LC would like to thank W. Buchm\"uller, S. Dittmaier, 
G. Moortgat-Pick and D. St\"ockinger for interesting and 
useful discussions.

The authors would like to acknowledge the hospitality of the CERN
Theory Division, where part of the project has been done. LC would
also like to thank the Physics Department of Lancaster University for
their kind hospitality, and MS would like to thank PPARC for a
studentship.

%%%%%%%%%%%%%%%%%%%%%%%%%%%%%%%%%%%%%%%%%%%%%%%%%%%%%%%%%%%%%%

\end{document}